\documentclass[aps,pra,multicolumn,twocolumn]{revtex4}
\usepackage{graphicx}
\usepackage{mathrsfs}
\usepackage{amssymb}
\usepackage{amsmath}

\begin{document}

\title{Engineering the coupling between Majorana bound states}
\author{Z. C. Shi$^{1,2}$, X. Q. Shao$^{3}$, Y. Xia$^{1,2,}$\footnote{xia-208@163.com}, and X. X. Yi$^{3,}$\footnote{yixx@nenu.edu.cn}}
\affiliation{$^1$ Department of Physics, Fuzhou University, Fuzhou 350002, China\\
$^2$ Fujian Key Laboratory of Quantum Information and Quantum Optics (Fuzhou University), Fuzhou 350116, China\\
$^3$ Center for Quantum Sciences and School of Physics, Northeast
Normal University, Changchun 130024, China}

\begin{abstract}
We study the coupling between Majorana bound states (CMBS), which is mediated by a topologically trivial chain in the presence of pairing coupling and long-range coupling. The
results show that CMBS can be enhanced by the pairing coupling and
long-range coupling of the trivial chain.  When driving the trivial chain by periodic driving field, we deduce the analytical expressions of CMBS in the high-frequency limit, and demonstrate that CMBS can be modulated by the frequency and amplitude of driving field. Finally we exhibit the application of tunable CMBS in realizing quantum logic gates.
\end{abstract}

 \maketitle

\section{introduction} \label{I}

Topological quantum computation \cite{kitaev03,nayak08}, immune to
certain types of noise, has attracted much attention since it was
proposed and becomes active again in recent years due to the
enormous progress in experiments. The gates used in topological quantum
computation is often conducted by creating quasi-particles, braiding
them, and measuring their states. For quasi-particles, two well-known types are Fibonacci anyons and Ising anyons [Majorana bound states (MBSs)]. The former are capable of offering universal
topological quantum computation, while the latter can  not form an
universal set of gates by only braiding operations. So
non-topologically protected gates have to be introduced in the MBSs-based
computation, which always requires coupling between
Majorana bound states (CMBS). Although it makes some breakthrough to pursue MBSs in
theories \cite{read00,ivanov01,fu08,hasan10,sau2010,qi11,alicea12,beenakker13}
and experiments \cite{das2012,mourik12,deng12,rokhinoson12,perge14,lee14}, the
question how to couple two MBSs is barely explored.

Recently, Schmidt and his co-workers \cite{schmidt13a,schmidt13}
presented proposals to  couple MBSs by putting the system into a
microwave cavity, where the microwave field can effectively
drive   population transfer between MBSs. The authors found that if
the microwave frequency approaches the band gap of the topologically
trivial region, CMBS is exponentially enhanced. In other words, CMBS is controllable by modulating microwave frequency or changing the number of photons in the cavity.
However, CMBS is relatively small in this system and one cannot very easily control the
parameters to realize unitary operation of qubit formed by MBSs.
In this work,  we study how to enhance CMBS by the pairing coupling and long-range coupling in the
central chain, and CMBS can be modulated by periodic driving field. In particular the
universal Majorana qubit rotation (UMQR) can be implemented by simply controlling the tunable
local gate voltage (TLGV) with square-wave form.

The paper is organized as follows. In Sec. \ref{II} we first briefly
review the Kitaev model and calculate CMBS in presence of pairing
coupling of the central chain. In Sec. \ref{III} we explore
the enhancement of CMBS when there exist pairing couplings at the
boundaries and long-range coupling in the trivial chain. In Sec. \ref{V} we
develop a scheme to control CMBS by modulating  the amplitude or the
frequency of driving field. Sec. \ref{VI} is devoted to
discussions and conclusions.

\section{CMBS induced by the trivial chain}   \label{II}

Consider an inhomogeneous Kitaev chain \cite{kitaev01} which can be
divided into three homogeneous parts. The total Hamiltonian of the whole chain reads
\begin{eqnarray}   \label{9}
H_{total}&=&H_{l}+H_{c}+H_{r}+H_{lc}+H_{rc},
\end{eqnarray}
where $H_{l}$ ($H_{r}$) denotes the Hamiltonian of left (right)
chain with sites from $-N_1$ ($N+1$) to $-1$ ($N_2$), and $H_{c}$
represents the Hamiltonian of central chain with sites from 0 to
$N$. $H_{lc}$ ($H_{rc}$) denotes the Hamiltonian of the coupling between the left
(right) chain and the central chain. To be specific the Hamiltonian of the three
homogeneous Kitaev chains is expressed as
\begin{eqnarray} \label{2a}
H_{\nu}&=&\sum_{n=M_{\nu}}^{M_{\nu}'}\mu_{\nu} a_{n}^{\dag}a_{n}
-\sum_{n=M_{\nu}}^{M_{\nu}'-1}(\frac{t_{\nu}}{2}a_{n}^{\dag}a_{n+1}      \nonumber\\
&&+\frac{\Delta_{\nu}e^{i\phi_{\nu}}}{2} a_{n}^{\dag}a_{n+1}^{\dag}+h.c.),
\end{eqnarray}
where $a_{n}$ and $a_{n}^{\dag}$ are the spinless fermion creation
and annihilation operators at site $n$ with  chemical potential
$\mu_{\nu}$ ($\nu=l,c,r$). $M_{\nu}=\{-N_1,0,N+1\}$ and
$M_{\nu}'=\{-1,N,N_2\}$ label the beginning and end sites of  the
left, central, and right chain, respectively. $t_{\nu}$ and
$\Delta_{\nu} e^{i\phi_{\nu}}$ are the hopping and pairing
amplitudes, respectively. Since the phase of the pairing amplitude
can be removed from the Hamiltonian by a gauge transformation, we
here and hereafter assume  both $t_{\nu}$ and $\Delta_{\nu}$   to be
real.

Physically, the Kitaev model can be exactly mapped into the
spin-$\frac{1}{2}$ chain with XY interactions \cite{lieb61}, the quantum system of the
semiconductor nanowire proximity coupling to a $s$-wave
superconductor \cite{oreg10,lutchyn10}, or the atomic chains \cite{nadjperge14,pawlak15}. This paradigm model shows
plentiful fascinating topological properties and there are two
distinct phases, i.e., the topologically trivial and nontrivial
phases. The critical points lie at $\frac{\mu_{\nu}}{t_{\nu}}=1$ and
$\Delta_{\nu}=0$ \cite{kitaev01}.

\begin{figure}[htbp]
\centering
\includegraphics[scale=0.45]{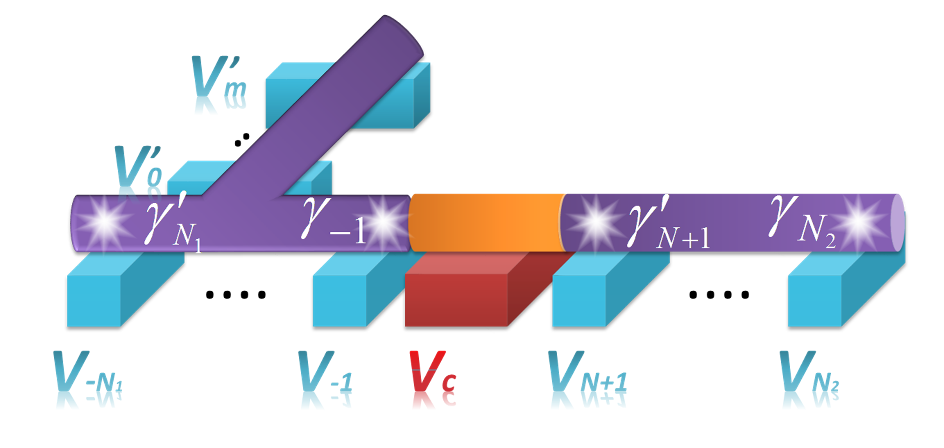}
\caption{Schematic illustration  for realizing universal Majorana
qubit rotation  in the Kitaev chain. There is  a direct relationship
between the Kitaev model and the solid state system in the
low-density limit (for details, see Refs.
\cite{alicea11,schmidt13a}). The braiding operation can be realized
in the T-junction on the left chain by adiabatically controlling
TLGV $V_i$ in sequence, and the CMBS can be modulated by the voltage $V_c$.}
\label{fig:02}
\end{figure}

By choosing $\mu_{l}=\mu_{r}=0$ (the chemical potential can be modulated by
TLGV $V_i$ shown in Fig. \ref{fig:02}) and
$\Delta_l=\Delta_r=t_l=t_r$ for the left and right chains,
both chains are in the topologically nontrivial phase.
As a result there exist MBSs at the ends of
both chains, depicted by the stars in Fig. \ref{fig:02}.
If one defines Majorana operators
\begin{eqnarray}  \label{3a}
\gamma_n=a_n+a_n^{\dag}, ~~\gamma_n'=i(a_n^{\dag}-a_n),~~n=-N_1,...,N_2,
\end{eqnarray}
after substituting them into Eq.(\ref{2a}), the Hamiltonian of left (right) chain becomes
\begin{eqnarray} \label{4a}
H_{\nu}=i\frac{t_{\nu}}{2}\sum_{n=M_{\nu}}^{M_{\nu}'-1}\gamma_n\gamma_{n+1}',~~~\nu=l,r.
\end{eqnarray}
Clearly, the Majorana operators $\gamma_n$ and $\gamma_{n+1}'$ are
coupled with different sites. In particular the Majorana operators
$\gamma_{N_1}'=i(a_{-N_1}^{\dag}-a_{-N_1})$,
$\gamma_{-1}=a_{-1}+a_{-1}^{\dag}$,
$\gamma_{N+1}'=i(a_{N+1}^{\dag}-a_{N+1})$, and
$\gamma_{N_2}=a_{N_2}+a_{N_2}^{\dag}$ do not appear in
Eq. (\ref{4a}), leading to the emergence of four MBSs in the chains.
Since the four MBSs are decoupled to the remaining sites and only
locate at the end of both left and right chains, we can ignore the
Hamiltonian of the remaining sites when studying CMBS.

\begin{figure}[htbp]
\centering
\includegraphics[scale=0.34]{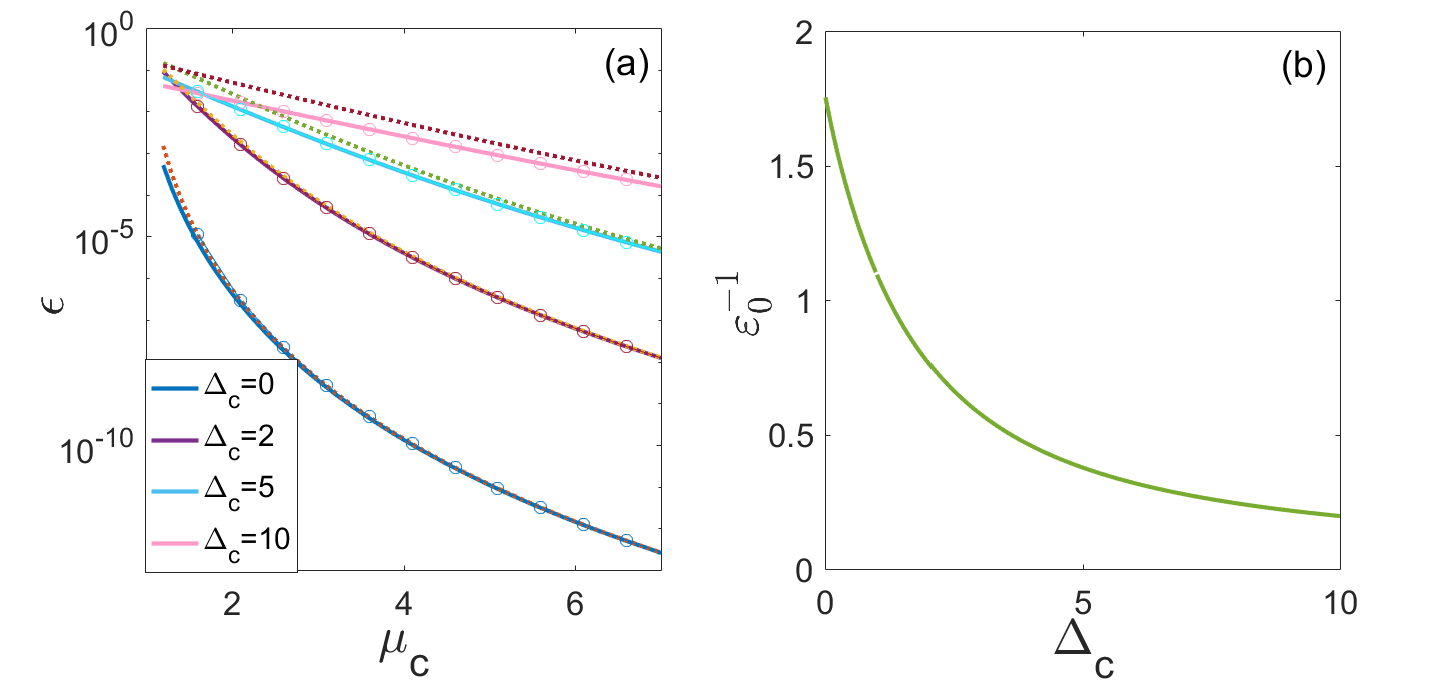}
\caption{(a) The coupling strength $\epsilon$ as a
function of the chemical potential $\mu_c$ with different pairing couplings $\Delta_c$, $N=10$.
The circle and solid lines are exact solutions by numerically diagonalizing the Hamiltonian (\ref{9}) with and without periodic boundary condition respectively, indicating no obvious difference by imposing periodic boundary condition. The dot line is approximate analytical
solutions described by Eq. (\ref{12}). All  parameters are in
units of hopping amplitude, i.e., $t_1=t_2=t_l=t_r=t_c=1$. (b) The coherence length versus the pairing amplitude with $\mu_c=3$.} \label{fig:03}
\end{figure}

To couple MBSs $\gamma_{-1}$ and $\gamma_{N+1}'$, the parameters $\{\mu_c,t_c,
\Delta_c\}$ should be chosen to guarantee the central chain is in
topologically trivial phase. Otherwise the MBSs $\gamma_{-1}$ and
$\gamma_{N+1}'$ would disappear in this inhomogeneous chain.
As we just consider the nearest-neighbour interactions in Eq. (\ref{2a}),
only the MBSs $\gamma_{-1}$ and
$\gamma_{N+1}'$ can be coupled to the central chain. So the
Hamiltonian of interest is given by \cite{bolech07}
\begin{eqnarray}      \label{10}
H_i&=&H_c+H_{lc}+H_{rc},                   \nonumber\\
H_{lc}&=&-\frac{t_{1}}{2}\gamma_{-1}(a_0-a_0^{\dag}),      \nonumber\\
H_{rc}&=&-\frac{it_{2}}{2}(a_N+a_N^{\dag})\gamma_{N+1}'.
\end{eqnarray}
In order to obtain the energy spectrum of the central chain,
it is convenient to carry out Fourier transform
$a_n=\frac{1}{\sqrt{N+1}}\sum_{k}a_{k}e^{ikn}$ by imposing periodic
boundary condition (we have set the lattice spacing to be unit and
Fourier transform will be precise when the number of sites is
large, $N\sim \infty$). The validity of Fourier transform is also
verified by numerical calculations in Fig. \ref{fig:03}.  The
Hamiltonian of the central chain now becomes
\begin{eqnarray}  \label{2}
H_c=\sum_{k}(\mu_c-t_c\cos k)a_{k}^{\dag}a_{k}-i\Delta_c\sin k a_{k}^{\dag}a_{-k}^{\dag}+h.c.
\end{eqnarray}
One can rewrite the above Hamiltonian in a normal Bogoliubov-de
Gennes (BdG) form by defining a two component operator
$A_k^{\dag}=[a_{k}^{\dag},a_{-k}]$, i.e.,
\begin{eqnarray}   \label{3}
H_c=\sum_{k}A_k^{\dag}\mathcal{H}_k A_k,~~~\mathcal{H}_k=\left(
                                                         \begin{array}{cc}
                                                           h_z & -ih_y \\
                                                           ih_y & -h_z \\
                                                         \end{array}
                                                       \right),
\end{eqnarray}
where $h_z=\mu_c-t_c\cos k$ and $h_y=\Delta_c\sin k$. Consequently
the Hamiltonian is further  simplified as (up to a constant)
\begin{eqnarray}  \label{4}
H_c=\sum_{k}E_{k}b_{k}^{\dag}b_{k},
\end{eqnarray}
where the quasi-particle operator $b_{k}$ is defined as
$b_{k}=u_{k}a_{k}-v_{k}a_{-k}^{\dag}$ and the energy spectrum of the
central chain is given by
$E_k=\sqrt{(\mu_c-t_c\cos{k})^2+\Delta_c^2\sin^2{k}}$,
$|u_k|=\sqrt{\frac{1}{2}(1+\frac{\mu_c-t_c\cos k}{E_k})}$,
$|v_k|=\sqrt{\frac{1}{2}(1-\frac{\mu_c-t_c\cos k}{E_k})}$. Using the
above notations, we can rewrite the Hamiltonian of interest
more explicitly as
\begin{eqnarray}  \label{11}
H_i&=&\sum_{k}E_{k}b_{k}^{\dag}b_{k}-\sum_{k}\frac{t_1}{2}\gamma_{-1}
[(u_k+v_k)b_{k}^{\dag}-(u_k^{*}+v_k^{*})b_{k}]
\nonumber\\ &&-\sum_{k}\frac{t_2}{2}[e^{-ikN}
(u_k-v_k)b_{k}^{\dag}+e^{ikN}(u_k^{*}-v_k^{*})b_{k}]\gamma_{N+1}'.   \nonumber\\
\end{eqnarray}

Since we are particularly interested in the low-energy
dynamics of system, we apply Schrieffer-Wolff transformation
\cite{schrieffer66} to  eliminate  high-energy spectrum. We denote
the first term in Eq. (\ref{11}) by $\mathcal{H}_0$ and the remaining
terms by $\mathcal{H}_1$. By choosing the unitary transformation
$\mathcal{S}$ to satisfy the relation
$[\mathcal{H}_0,\mathcal{S}]=\mathcal{H}_1$, one has
\begin{eqnarray}  \label{11a}
&&\mathcal{S}=\sum_{k}\frac{t_1(u_k+v_k)\gamma_{-1}b_{k}-t_1(u_k^{*}+v_k^{*})b_{k}^{\dag}\gamma_{-1}}{2 E_k}    \nonumber\\
&&+\sum_{k}\frac{t_2e^{ikN}(u_k-v_k)\gamma_{N+1}'b_{k}-t_2e^{-ikN}
(u_k^{*}-v_k^{*})b_{k}^{\dag}\gamma_{N+1}'}{2 E_k}.  \nonumber\\
\end{eqnarray}
Making use of the Baker-Hausdorff formula:
$e^{-\mathcal{S}}H_ie^{\mathcal{S}}=H_i+[H_i,\mathcal{S}]+\frac{1}{2} [[H_i,\mathcal{S}],\mathcal{S}]+\cdots$,
and keeping the terms up to first order, we obtain an effective Hamiltonian between the adjacent MBSs $\gamma_{-1}$
and $\gamma_{N+1}'$, i.e.,
\begin{eqnarray}     \label{12}
H_{0N}=i\epsilon\gamma_{-1}\gamma_{N+1}', ~~~\epsilon=
\frac{t_1t_2}{\sqrt{\mu_c^{2}+\Delta_c^2-t_c^2}}e^{-N/\varepsilon_0}.
\end{eqnarray}
Here $\varepsilon_0$ is the coherence length,
$\varepsilon_0^{-1}=\ln\frac{\Delta_{c}-t_c}{\sqrt{\mu_c^{2}+\Delta_c^2-t_c^2}-\mu_c}$. $\epsilon$ represents the amplitude for the MBSs tunneling across the central chain.
Eq. (\ref{12}) demonstrates that the effective coupling between MBSs $\gamma_{-1}$ and
$\gamma_{N+1}'$ can be induced by the central chain. This is a
virtual co-tunneling process since there exists energy gap in the
central chain and real electrons and holes cannot tunnel from one
MBS to another. In addition, the coupling strength $\epsilon$ depends
exponentially on the length $N$. When $N$ is large enough, the degeneracy energies of MBSs are in the energy gap, which is the reason why MBSs are topologically protected.
In Fig. \ref{fig:03}(a),  we plot the
coupling strength $\epsilon$ versus the chemical potential $\mu_c$
with different pairing amplitudes $\Delta_c$.
It is readily observed that the analytical solutions of $\epsilon$ are in well agreement with the
numerical solutions when the chemical potential is large enough and the coupling
strength $\epsilon$ decreases with the increasing of chemical
potential $\mu_c$, reminiscent of the results where
the Majorana qubit setup is placed in a microwave cavity
\cite{schmidt13a}. However, one also finds that the coupling
strength $\epsilon$ is enhanced by the pairing amplitude
$\Delta_c$ of the central chain, which is not mentioned in Ref. \cite{schmidt13a}.
Physically, it originates from the fact
that large pairing amplitude $\Delta_c$ would broaden the energy spectrum [cf. $E_k$ in Eq. (\ref{4})], leading to more quasi-particles participating
in the co-tunneling process. As a result the coherence length
$\varepsilon_0$ also increases accordingly, which is verified in Fig. \ref{fig:03}(b).

\section{the enhancement of CMBS}  \label{III}

In Eq. (\ref{10}), we just consider the hopping coupling at the boundaries between left (right) chain and central chain. When existing pairing coupling at the boundaries, the Hamiltonian of interest becomes
\begin{eqnarray}
H_i&=&H_c+H_{lc}+H_{rc},                   \nonumber\\
H_{lc}&=&-\frac{t_{1}}{2}\gamma_{-1}(a_0-a_0^{\dag})-
\frac{\Delta_1}{2}(a_{-1}^{\dag}a_{0}^{\dag}-a_{-1}a_{0}),      \nonumber\\
H_{rc}&=&-\frac{it_{2}}{2}(a_N+a_N^{\dag})\gamma_{N+1}'-
\frac{\Delta_2}{2}(a_{N}^{\dag}a_{N+1}^{\dag}-a_{N}a_{N+1}), \nonumber\\
\end{eqnarray}
where $H_c$ is the same as in Eq. (\ref{10}).
By using the Majorana operators representation in Eq. (\ref{3a}), the
Hamiltonian can be rewritten as,
\begin{eqnarray}   \label{14}
H_{lc}&=&-\frac{t_{1}+\Delta_1}{2}\gamma_{-1}(a_0-a_0^{\dag}) +\frac{i\Delta_1}{2}\gamma_{-1}'(a_0+a_0^{\dag}),      \nonumber\\
H_{rc}&=&-\frac{i(t_{2}+\Delta_2)}{2}(a_N+a_N^{\dag})\gamma_{N+1}' +\frac{\Delta_2}{2}(a_N-a_N^{\dag})\gamma_{N+1}. \nonumber\\
\end{eqnarray}
The second term in the Hamiltonian $H_{lc}$ $(H_{rc})$ can be
ignored if we only consider CMBS, but  this term affects the spatial
distribution of the MBSs $\gamma_{-1}$ and $\gamma_{N+1}'$. Hence
the effective hopping coupling between the central chain and
the left (right) chain is enhanced as $\frac{t_1+\Delta_1}{2}(\frac{t_2+\Delta_2}{2})$.
As a result, the value of coupling strength $\epsilon$ increases in the existence of pairing coupling at the boundaries.

\begin{figure}[htbp]
\centering
\includegraphics[scale=0.3]{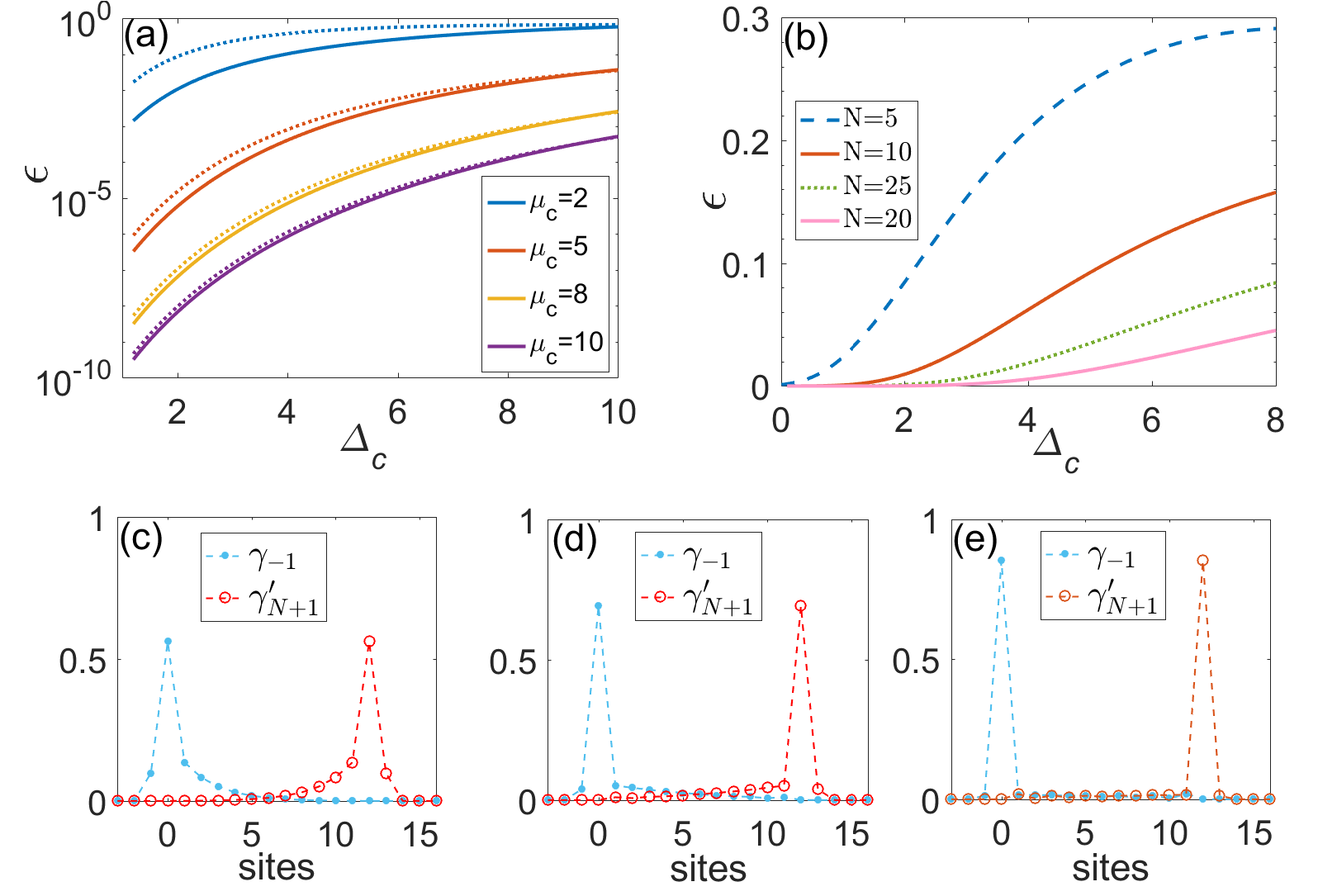}
\caption{(a) The coupling strength $\epsilon$ as a
function of the pairing amplitude $\Delta_c$
with different chemical potentials $\mu_c$, $N=10$. The solid
line and dot line denote the exact numerical solutions and the approximate
analytical solutions respectively, where the hopping amplitude $t_1t_2$ are
revised as $(t_1+\Delta_1)(t_2+\Delta_2)$. All
 parameters are chosen in units of hopping amplitude in the central chain.
$\Delta_1=\Delta_2=5.$  (b) The exact
numerical solutions of coupling strength $\epsilon$ versus the pairing amplitude $\Delta_c$ when
$\mu_c=2$, $\Delta_1=\Delta_2=1$. (c)-(e) The spatial density distribution of MBSs $\gamma_{-1}$ and
$\gamma_{N+1}'$ versus different coupling strength $\epsilon$. (c) $\epsilon=0.0337$. (d) $\epsilon=0.2055$. (e) $\epsilon=0.2519$.  }  \label{fig:04}
\end{figure}

Figs. \ref{fig:04}(a)-(b) demonstrate the coupling strength $\epsilon$  as
a function of the pairing amplitude $\Delta_c$. It is
observed that the analytical solutions are in well agreement with the
numerical solutions when $\mu_c$ is large enough, and the coupling
strength $\epsilon$ can reach a relatively high value even though
the central chain is long.
Note that CMBS would in turn affect the
spatial distribution of MBSs $\gamma_{-1}$ and $\gamma_{N+1}'$.
Especially, the spatial distribution of MBSs will be greatly modified when the coupling strength $\epsilon$ is large, as shown in Figs. \ref{fig:04}(c)-(e).

\begin{figure}[htbp]
\centering
\includegraphics[scale=0.45]{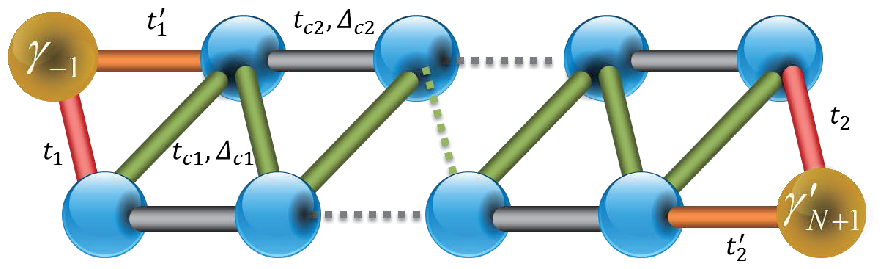}
\caption{The zigzag structure of the Kitaev chain which can  be
realized  in  optical lattices. }  \label{fig:05}
\end{figure}

On the other hand, when the central chain exists long-range coupling [e.g., the next-nearest-neighbor (NNN) couplings], the Hamiltonian can be described by
\begin{eqnarray}
H_c&=&\sum_{n=0}^{N}\mu_{c} a_{n}^{\dag}a_{n}-\sum_{m=1}^{2}\sum_{n=0}^{N-m}(\frac{t_{cm}}{2}a_{n}^{\dag}a_{n+m}  \nonumber\\
&&+\frac{\Delta_{cm}}{2} a_{n}^{\dag}a_{n+m}^{\dag}+h.c.).
\end{eqnarray}
Physically, with the help of Raman laser, this model can
be realized by the system of fermi atoms trapping in optical lattice
with zigzag structure and coupling to a 3D Bose-Einstein condensate (BEC) reservoir, as shown
in Fig. \ref{fig:05}. The relative strength of hopping amplitudes
$t_{c1}$ and $t_{c2}$ can be modulated by changing the zigzag
geometry. For details, we refer readers to Ref. \cite{jiang11,kraus12}.
In this case, the coupling Hamiltonian at the boundaries between left (right) chain and central chain reads
\begin{eqnarray}
H_{lc}&=&-\frac{t_{1}}{2}\gamma_{-1}(a_0-a_0^{\dag})-\frac{t_{1}'}{2}\gamma_{-1}(a_1-a_1^{\dag}),      \nonumber\\
H_{rc}&=&-\frac{it_{2}}{2}(a_N+a_N^{\dag})\gamma_{N+1}'-\frac{it_{2}'}{2}(a_{N-1}+a_{N-1}^{\dag})\gamma_{N+1}'. \nonumber\\
\end{eqnarray}

\begin{figure}[htbp]
\centering
\includegraphics[scale=0.34]{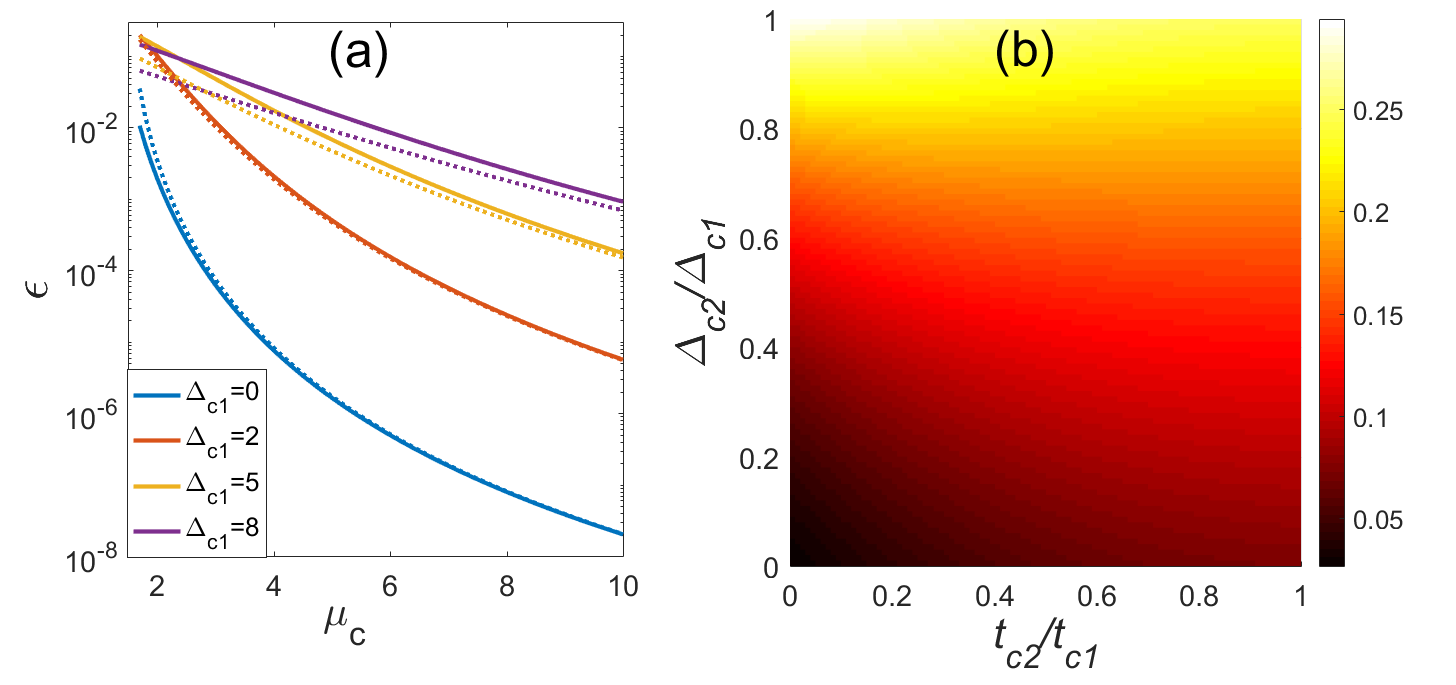}
\caption{ (a) The coupling strength $\epsilon$
as a function of the chemical potential $\mu_c$.
$N=10$. $t_{c2}=\frac{1}{2}t_{c1}$.
$\Delta_{c2}=\frac{1}{2}\Delta_{c1}$.
$t_1=t_2=t_{c1}$. $t_1'=t_2'=t_{c2}$. The solid line and dot line
denote the exact numerical solutions and the approximate analytical
solutions, respectively. (b) The coupling strength $\epsilon$
as a function of the hopping amplitude and pairing
amplitude. $\mu_c=2$. $\Delta_{c1}=5$.  }  \label{fig:06}
\end{figure}

Following the similar derivation procedures  in
Eqs. (\ref{10})-(\ref{12}), one can estimate the effective
Hamiltonian of CMBS,
\begin{eqnarray}
H_{0N}'&=&i\epsilon\gamma_{-1}\gamma_{N+1}'     \nonumber\\
\epsilon&=&\sum_{k}\Big{|}\lim_{z\rightarrow z_k}(z-z_k)\frac{2(t_1z+t_1')(t_2z+t_2')z^{N-1}}{\prod_{i=1}^{4}(z-z_{i})}\Big{|}, \nonumber\\
z_k&\in&\Big{\{}z_i\Big{|}|z_i|<1,~i=1,2,3,4\Big{\}},
\end{eqnarray}
where $z_i$ ($i=1,2,3,4$) is the root of the quartic equation
$(\Delta_{c2}-t_{c2})z^4+(\Delta_{c1}-t_{c1})z^3+2\mu_cz^2-(\Delta_{c1}+t_{c1})z-(\Delta_{c2}+t_{c2})=0$.
Obviously, the Hamiltonian returns   to Eq. (\ref{12}) if
one sets $t_{c2}=\Delta_{c2}=0$. Fig. \ref{fig:06} depicts the
relation between the coupling strength and the parameters of the
central chain. One observes that the analytical solutions are in
well agreement with the numerical solutions when $\mu_c\gg t_c$,
i.e., $E_k\gg t_{1,2}$. In the presence  of long-range coupling,
the coupling strength $\epsilon$ is also enhanced since it takes  minimum when
$t_{c2}=\Delta_{c2}=0$, as shown in Fig. \ref{fig:06}(b). The results
are not surprising since there are multiple channels for electrons
co-tunneling in the long-range coupling regime, i.e., the
NNN hopping amplitude also makes contributions to the
electrons co-tunneling process.

\section{Modulation of CMBS by periodic driving}   \label{V}

As shown in Sec. \ref{II}, CMBS is closely related to the chemical potential,
the hopping amplitude, and the pairing amplitude of the central
chain [cf. Eq. (\ref{12})], which implies that we can modulate the coupling
strength $\epsilon$ by varying the values of those parameters. Considering the hopping amplitude and the pairing amplitude cannot be easily manipulated in practice, we
first explore the relation between the coupling strength
$\epsilon$ and the chemical potential $\mu_c$. Here, we do not intend to plot
directly the dependence of $\epsilon$ on $\mu_c$. Instead,
we show the dependence of Rabi oscillation between two MBSs on
$\mu_c$. By revealing different values of Rabi frequency, it can show not only how the coupling
strength changes with the chemical potential [since distinct
coupling strengths can reflect on distinct Rabi frequencies, cf.
Eq. (\ref{21})], but also how well the Majorana qubit works.

\begin{figure}[htbp]
\centering
\includegraphics[scale=0.4]{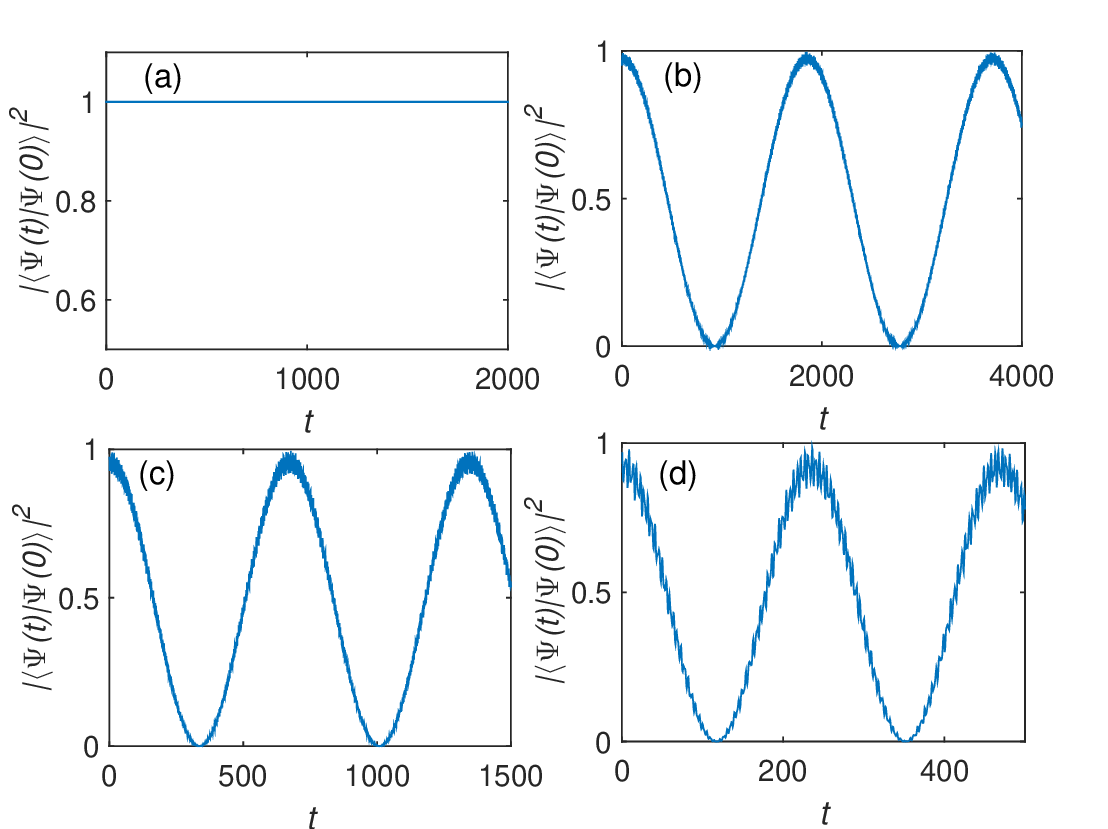}
\caption{ Population of $|\Psi(0)\rangle$ as a function of evolution
time. The system is in the MBS $\gamma_{-1}$ initially. The other
parameters are $\mathcal{N}=20$, $\mu_l=\mu_r=0$,
$\Delta_l=\Delta_r=t_l=t_r=5$,
$\Delta_c=5$, $t_1=t_2=t_c$, $N=10$. (a) $\mu_c=10$.
(b) $\mu_c=3$. (c) $\mu_c=2.5$. (d) $\mu_c=2$. }  \label{fig:07}
\end{figure}

\begin{figure}[htbp]
\centering
\includegraphics[scale=0.45]{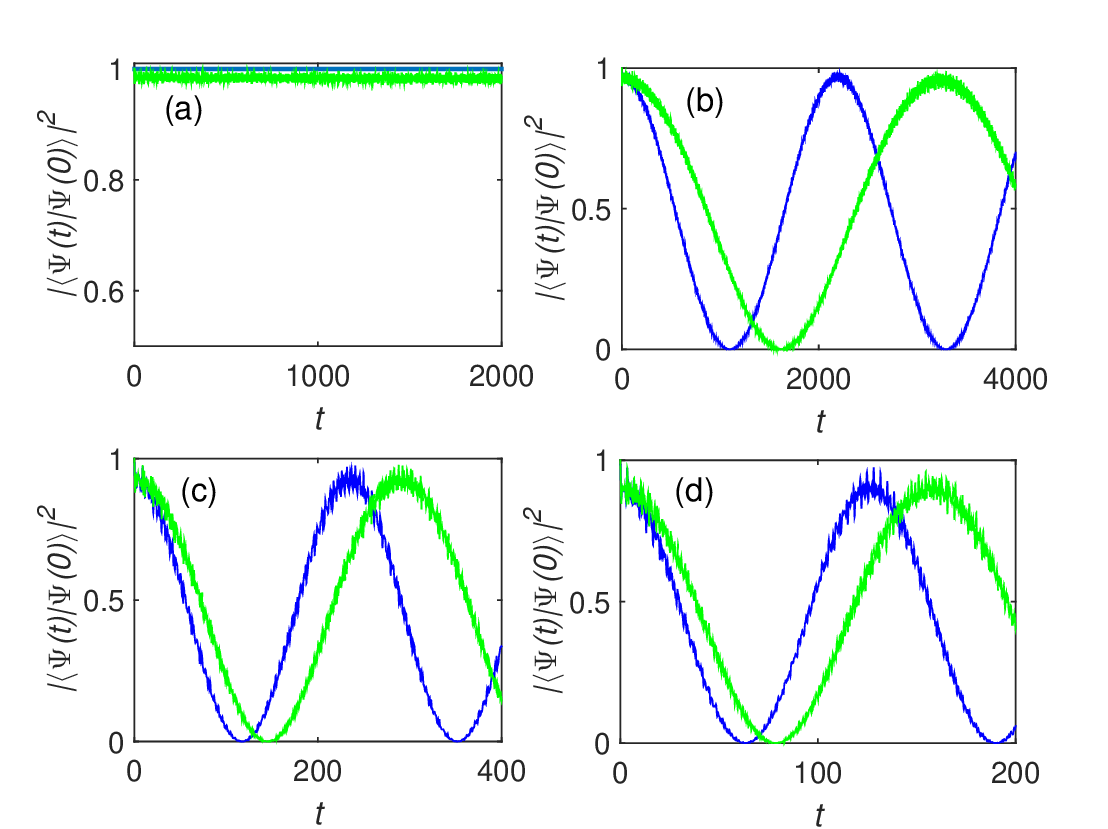}
\caption{ Population of $|\Psi(0)\rangle$ as a function of evolution
time. The system is in the MBS $\gamma_{-1}$ initially. The other
parameters are $\mu_l=\mu_r=0$, $\Delta_l=\Delta_r=t_l=t_r=5$,
$\mu_c=2$, $t_1=t_2=t_c$, $N=10$. (a) $\Delta_c=0$.  (b)
$\Delta_c=2$. (c) $\Delta_c=5$. (d) $\Delta_c=10$. The blue-dark
lines are the exact numerical results obtained by Eq. (\ref{9}), and
the green-grey lines are plotted  by the Hamiltonian given
in Eq. (\ref{19}). The effective pairing amplitude are the same for
two lines in each panels.} \label{fig:08}
\end{figure}

To demonstrate Rabi oscillation, we first write down the total Hamiltonian
 of inhomogeneous chain in the BdG form:
$H_{total}=\frac{1}{2}A^{\dag}\mathcal{H}_{total}A$, where the basis
$A=[a_{-N_{1}},...,a_{N_{2}}, a_{-N_{1}}^{\dag},...,a_{N_{2}}^{\dag}]^{T}$.
In fact, the MBSs is the eigenstates of BdG Hamiltonian
$\mathcal{H}_{total}$ corresponding to zero eigenvalue.
Suppose that the inhomogeneous chain is initially prepared in the left MBS $\gamma_{-1}$, i.e., $|\Psi(0)\rangle=\frac{1}{\sqrt{2}}(|-1\rangle+|\mathcal{N}\rangle)$.
Here, the label $|m\rangle$ denotes the vector
with components $|m\rangle_j=\delta_{mj}$ in the basis $A$,
$m,j\in\{-N_1,...,2\mathcal{N}\}$, and $\mathcal{N}$ denotes
the total number of sites.
With these notations, the dynamics of inhomogeneous chain is governed by following equation \cite{gennes66}
\begin{eqnarray}
i\frac{d|\Psi(t)\rangle}{dt}=\mathcal{H}_{total}|\Psi(t)\rangle,
\end{eqnarray}
In Fig. \ref{fig:07}, we plot the population of $|\Psi(0)\rangle$ (i.e., the left MBS $\gamma_{-1}$) as a function of evolution time. If there does not exist CMBS, the Rabi oscillation can not appear  in the
system, as depicted in Fig. \ref{fig:07}(a). In Figs.
\ref{fig:07}(b)-(d), different chemical potentials $\mu_c$ result in
 different Rabi frequencies. When the chemical potential is large,
the coupling strength would be small, leading to a small Rabi
frequency. That is, the Rabi frequency increases with the
decreasing of the chemical potential. Similarly, one can modulate the coupling
strength by changing the pairing amplitude, as
the blue-dark lines show in Fig. \ref{fig:08}. Besides, the coupling
strength can also be modulated purely by the phase of pairing amplitude,
e.g., Refs. \cite{alicea11,zyuzin13,tong13,xue14}.

Since the pairing amplitude is inherently determined by the property
of superconductors, it may be difficult to directly modulate with current techniques. In following we show
that this goal can be reached by periodically driving the central chain, and the Hamiltonian of periodic driving field reads
\begin{eqnarray}
H_{\mu}(t)=\mu_{0}\cos{\omega t}\sum_{n=0}^{N}a_{n}^{\dag}a_{n},
\end{eqnarray}
where $\mu_{0}$ and $\omega$ are the amplitude and the frequency of the periodic
driving field, respectively. In a realistic situation, this
driving field  can be achieved by applying an external
ac electric potential to TLGV, since the on-site chemical potential
can be modulated by TLGV.

\begin{figure}[htbp]
\centering
\includegraphics[scale=0.45]{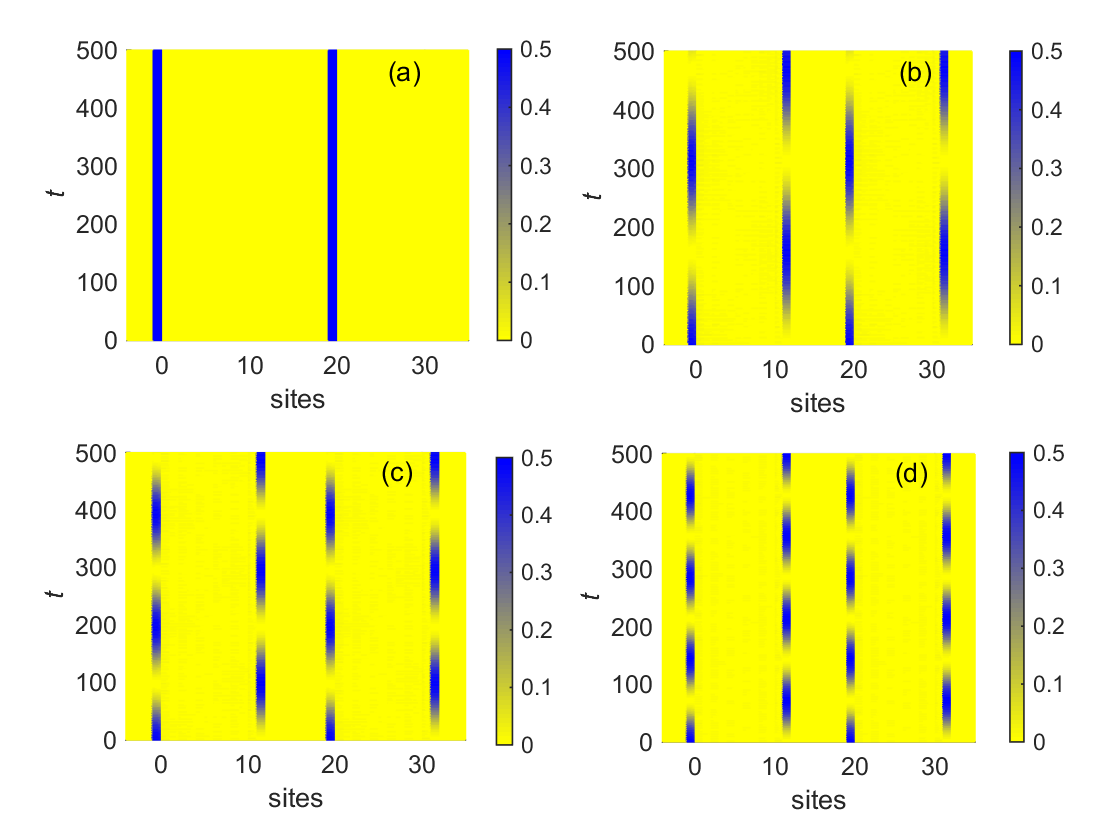}
\caption{ The population of each sites in $|\Psi\rangle$ as a
function of time. The MBS $\gamma_{-1}$ locates at the site $-1$
initially. $\Delta_c=10$. $N=20$. (a)
$\mathcal{J}_{0}(\frac{\mu_0}{\omega})=0$. (b)
$\mathcal{J}_{0}(\frac{\mu_0}{\omega})=0.8605$. (c)
$\mathcal{J}_{0}(\frac{\mu_0}{\omega})=0.9120$. (d)
$\mathcal{J}_{0}(\frac{\mu_0}{\omega})=0.9696$. The other physical
parameters are the same as in Fig. \ref{fig:08}. The three part of
Kitaev chain are decoupled from each other when
$\mathcal{J}_{0}(\frac{\mu_0}{\omega})=0$, and there exists Rabi
oscillation between the adjacent MBSs $\gamma_{-1}$ and $\gamma_{N+1}'$ when
$\mathcal{J}_{0}(\frac{\mu_0}{\omega})\neq 0$, where the MBSs
$\gamma_{N+1}'$ locates at site $11$.} \label{fig:09}
\end{figure}

At first, in the rotation frame defined by the unitary transformation,
$U(t)=e^{-i\frac{\mu_{0}}{\omega}\sin{\omega
t}\sum_{n=0}^{N}a_{n}^{\dag}a_{n}}$, the effective Hamiltonian
of the whole chain reads,
\begin{eqnarray} \label{19}
H_{total}'&=&U^{\dag}(t)[H_{total}+H_{\mu}(t)]U(t)-iU^{\dag}(t)\dot{U}(t),        \nonumber\\
&=&H_{l}+H_{r}+H_{c}'+H_{lc}'+H_{rc}',    \nonumber\\
H_{c}'&=&\sum_{n=0}^{N}\mu_{c} a_{n}^{\dag}a_{n}-\sum_{n=0}^{N-1}(\frac{t_{c}}{2}a_{n}^{\dag}a_{n+1}            \nonumber\\
&&+ e^{i\frac{2\mu_0}{\omega}\sin(\omega t)}\frac{\Delta_{c}}{2} a_{n}^{\dag}a_{n+1}^{\dag}+h.c.),    \nonumber\\
H_{lc}'&=&-e^{-i\frac{\mu_0}{\omega}\sin(\omega t)}\frac{t_1}{2}a_{-1}^{\dag}a_0+h.c.,    \nonumber\\
H_{rc}'&=&-e^{-i\frac{\mu_0}{\omega}\sin(\omega t)}\frac{t_2}{2}a_{N+1}^{\dag}a_N+h.c.,
\end{eqnarray}
where $H_l$ and $H_r$ are invariant under this rotation.
By making use of the identity
\begin{eqnarray}
e^{ix\sin\omega t}=
\sum_{n=-\infty}^{\infty}\mathcal{J}_{n}(x)e^{i
n\omega t},
\end{eqnarray}
where $\mathcal{J}_{n}(x)$ is the $n$-order Bessel function, the
time-dependent effective Hamiltonian in the high-frequency limit
(i.e., $\omega\gg \mu_{c}, t_{c}$) becomes
\begin{eqnarray}  \label{22}
H_{c}'&=&\sum_{n=0}^{N}\mu_{c} a_{n}^{\dag}a_{n}-\sum_{n=0}^{N-1}(\frac{t_{c}}{2}a_{n}^{\dag}a_{n+1}  \nonumber\\
&&+\frac{\Delta_{c}}{2}\mathcal{J}_{0}(\frac{2\mu_0}{\omega}) a_{n}^{\dag}a_{n+1}^{\dag}+h.c.),   \nonumber\\
H_{lc}'&=&-\frac{t_1}{2}\mathcal{J}_{0}(\frac{\mu_0}{\omega})a_{-1}^{\dag}a_0+h.c.,    \nonumber\\
H_{rc}'&=&-\frac{t_2}{2}\mathcal{J}_{0}(\frac{\mu_0}{\omega})a_{N+1}^{\dag}a_N+h.c.
\end{eqnarray}
One finds from Eq. (\ref{22}) that the pairing amplitude of central chain
is modulated by the amplitude and the frequency of driving field
through zero-order Bessel function, i.e.,
$\Delta_{eff}=\frac{\Delta_{c}}{2}\mathcal{J}_{0}(\frac{2\mu_0}{\omega})$.
In the presence of driving field, we plot the time evolution of the MBSs by the green-grey lines in Fig. \ref{fig:08}. The results show that, even though the effective pairing amplitude with driving field equals to the pairing amplitude without driving field, the Rabi
frequency is still a bit different in two cases. This originates from the fact that the effective hopping amplitudes at boundaries are also changed by driving
field [see the expressions of $H_{lc}'$ and $H_{rc}'$ in Eq. (\ref{22})], rendering the
correction of Rabi frequency. We would like to address that, this
correction does not make difference for UMQR as it only changes evolution time to complete corresponding operations. Interestingly, we can also modulate
the effective hopping amplitude at boundaries by driving
field to control CMBS. It is demonstrated in Fig. \ref{fig:09}(a)
that the adjacent MBSs $\gamma_{-1}$ and $\gamma_{N+1}'$ are
decoupled by making the effective hopping amplitude at boundaries
($H_{lc}'$ and $H_{rc}'$) vanish, i.e., setting
$\mathcal{J}_{0}(\frac{\mu_0}{\omega})=0$. Otherwise, CMBS can be
really induced, manifesting in the Rabi oscillation shown
in Figs. \ref{fig:09}(b)-(d).

\section{discussion and conclusion}  \label{VI}

As well known, an ordinary spinless fermion can be used to encode a
logical qubit because it can span a two-dimensional Hilbert
space (occupy or empty). However it is not true for MBSs
since the operators satisfy $\gamma_{i}=\gamma_{i}^{\dag}$ and
$\gamma_{i}^2=1$. By recombining the operators $\gamma_{i}$, one can
use two MBSs to construct a Dirac  fermion, e.g.,
$d_1=\frac{1}{2}(\gamma_{N_1}'+i\gamma_{-1})$ and
$d_2=\frac{1}{2}(\gamma_{N+1}'+i\gamma_{N_2})$ in Fig. \ref{fig:02}.
It seems that a logical qubit can be encoded by two MBSs now.
Nevertheless, for the system with parity conservation
(calculated through the Dirac fermions formed by MBSs), the coherent
superposition of  MBSs with  different parities is prohibited.
Therefore a logical qubit cannot be encoded by two MBSs in the parity
conservation system. To guarantee two computational bases
having same parity, four MBSs is necessary (see $\gamma_{N_1}'$,
$\gamma_{-1}$, $\gamma_{N+1}'$, and $\gamma_{N_2}$ in Fig.
\ref{fig:02}). For instance, we
can construct the Majonara-based qubit in the odd parity subspace,
\begin{eqnarray}
|1_{1}0_{2}\rangle=d_1^{\dag}|0_{1}0_{2}\rangle, |0_{1}1_{2}\rangle=d_2^{\dag}|0_{1}0_{2}\rangle,
\end{eqnarray}
where $|0_{1}0_{2}\rangle$ is the vacuum state of Dirac
fermions. The topologically protected single qubit operations are
achieved by  exchanging spatial positions of the MBSs $\gamma_{N_1}'$
and $\gamma_{-1}$, which exhibit the non-Abelian statistics. This
braiding operation, i.e., the $\frac{\pi}{4}$
phase gate, is represented by the following unitary operation
\begin{eqnarray}
U_{N_10}=e^{\frac{i\pi}{4}\sigma_{z}},
\end{eqnarray}
where $\sigma_z=|1_{1}0_{2}\rangle\langle
1_{1}0_{2}|-|0_{1}1_{2}\rangle\langle 0_{1}1_{2}|$. This process
can be realized in the one-dimensional semiconducting
wires with T-junction by controlling TLGV adiabatically
\cite{alicea11}. Note that the braiding operation is insufficient
for realizing universal quantum computation since
it is not able to perform arbitrary
single qubit rotations \cite{nayak08,sravyi06}, which are
usually not topologically protected (rotation angle
$\theta\neq \frac{\pi}{2}n$, $n$ is integer).

When there exists CMBS in the system (it has been also investigated
in the continuous model instead of lattice model
\cite{cheng09,sau11a,sarma12,prada12,rainis13,cottet13,kovalev14}), the
effective Hamiltonian takes
$H_{0N}=\frac{i\epsilon}{2}\gamma_{-1}\gamma_{N+1}'$. We can obtain
the following unitary operation for a fixed evolution time
\begin{eqnarray}   \label{21}
U_{0N}(t)=e^{-\frac{i\epsilon t}{2}\sigma_x},
\end{eqnarray}
where $\sigma_x=|1_{1}0_{2}\rangle\langle
0_{1}1_{2}|+|0_{1}1_{2}\rangle\langle 1_{1}0_{2}|$. Together with
the braiding operation $U_{N_10}$, one can implement UMQR via the
successive operations $U'=U_{0N}(t_2)U_{N_10}U_{0N}(t_1)$
\cite{schmidt13a}. To realize the operation $U'$, exact
control over CMBS with a fixed evolution time is required. This is
crucial since the procedure is usually not topologically protected.
The other consideration is that the MBSs should decouple
instantaneously from each other after (before) the coupling of the
adjacent MBSs. As shown in Figs. (\ref{fig:07})-(\ref{fig:09}), the
two considerations are sufficiently solved since the modulation of
the amplitude or frequency of the electric potential on TLGV can be
controlled with high precision. For example, in order to
manipulate the chemical potential, we adopt two distinct voltages,
e.g., $\mu_{c1}=10$ and $\mu_{c2}=2.5$ as shown in  Fig.
\ref{fig:07}(a) and Fig. \ref{fig:07}(c). We first apply  a low
voltage $\mu_{c2}$ for periods of time $t_1$ to realize the
operation $U_{0N}(t_1)$. Then  we switch   it to a high voltage
$\mu_{c1}$ to realize the braiding operation $U_{N_10}$. Next we
change the gate  to the low voltage $\mu_{c2}$ for periods of time
$t_2$ to realize the operation $U_{0N}(t_2)$. Finally we take back to
the high voltage $\mu_{c1}$ to cancel CMBS. This suggests that
UMQR can be realized by changing TLGV in a square-wave form,
composed by $\mu_{c1}$ and $\mu_{c2}$. In fact, the results are the
same as in the case where the driving frequency is switched to
$\omega_1$ or $\omega_2$ in sequence; see Fig. \ref{fig:09}(a) and
Fig. \ref{fig:09}(c).

In conclusion, we have studied CMBS in the inhomogeneous Kitaev chain with short-range and long-range coupling.
We demonstrate that CMBS depends sharply on the pairing amplitude
and the chemical potential of the central chain.
Particularly, CMBS is remarkably enhanced when existing the pairing
amplitude or long-range coupling. In additional we have explored the dependence of CMBS on the
frequency and amplitude of the driving field. These results suggest
that there are many ways to change CMBS such as manipulating
the chemical potential of central chain, or
modulating the effective pairing amplitude of central chain
which can be realized by changing the frequency (or amplitude) of driving
field. Finally, we have demonstrated the application of tunable CMBS, i.e., implementing
unitary rotation operations for quantum computation.

\section*{ACKNOWLEDGMENTS}

This work is supported by the National Natural Science Foundation
of China (Grant Nos. 11575045, 11374054, 11674060, 11534002, and 61475033), the Major State Basic Research Development Program of China under Grant No. 2012CB921601, and the fund from Fuzhou University under No. XRC-1639.

\end{document}